# Photoionization, fluorescence, and inner-shell processes

Much of the light (photons) from stars ultimately originates from the photon emission from a particular atom or molecule at an exact wavelength (color). These wavelengths allow us to uniquely identify the atoms and molecules that distant stars are composed of without the considerable effort required to travel there. For example, closer to home, the vivid green and blue colors observed in the aurora borealis (the northern lights) correspond to emissions from oxygen and nitrogen, respectively. Most of the known matter in the universe is in a plasma state (the state of matter similar to gas in which a certain portion of the particles are ionized) and our information about the universe is carried by photons (light), which are dispersed and detected for example by the orbiting *Chandra X-ray Observatory*. When photons travel through stellar atmospheres and nebulae (interstellar clouds of gas and dust), they are likely to interact with matter and therefore with ions. This makes the study of photoionization (the physical process in which an incident photon ejects one or more electrons from an atomic or molecular system) of atoms, molecules, and their positive ions very important for astrophysicists, helping them to interpret stellar data (spectroscopy). To measure the chemical evolution of the universe and then understand its ramifications for the formation and evolution of galaxies and other structures is a major goal of astrophysics today. The answers to these questions ultimately address human and cosmic origins. Our ability to infer chemical abundances relies extensively on spectroscopic observations of a variety of low-density cosmic plasmas including the diffuse interstellar and intergalactic media, gas in the vicinity of stars, gas in supernova remnants, and gas in the nuclei of active galaxies. Cosmic chemical evolution is revealed through emission or absorption lines from cosmically abundant elements.

The quantitative information we have about the cosmos is the result of spectroscopy from the ground and an array of orbital missions: *International Ultraviolet Explorer* (*IUE*); *Hubble Space Telescope* (*HST*); *Extreme Ultraviolet Explorer* (*EUVE*); *Infrared Space Observatory* (*ISO*); Hopkins Ultraviolet Telescope (HUT); *Orbiting and Retrievable Far and Extreme Ultraviolet Spectrograph-Shuttle Pallet Satellite* (*ORFEUS-SPAS*); *Solar and Heliospheric Observatory* (*SOHO*); *Far-Ultraviolet Spectrograph Explorer* (*FUSE*); *Chandra X-ray Observatory*; *XMM-Newton*; *Galaxy Evolution Explorer* (*GALEX*); *Constellation-X*, now the *International X-ray Observatory* (*IXO*); *Suzaku*, a reflight of *ASTRO-E*; and soon, *ASTRO-H*.

Satellites launched in 1987 indicated that the x-ray spectra of active galactic nuclei (AGN) were not featureless continua, but contained structure associated with various atomic processes. Iron emission lines from inner shells at wavelengths of 0.1936 and 0.1940 nm (1 nm = $10^{-9}$ m) were observed. These lines are characteristic of x-ray-illuminated cold material, and act as an indication of the amount of cool material illuminated by the AGN. X-ray satellites, such as *Chandra* and *XMM-Newton*, launched over a decade ago, have provided for the first time the ability to do x-ray spectroscopy at a very high resolution. Elements of prime interest are carbon, nitrogen, oxygen, neon, magnesium, silicon, sulphur, calcium, argon, and iron, primarily at the *K*- and *L*-edge energies, 0.2–12 keV (**Fig, 1**). The *Chandra* and *XMM-Newton* satellites currently provide a wealth of spectral data, which spectral modeling codes struggle to interpret due to a lack of quality atomic data.

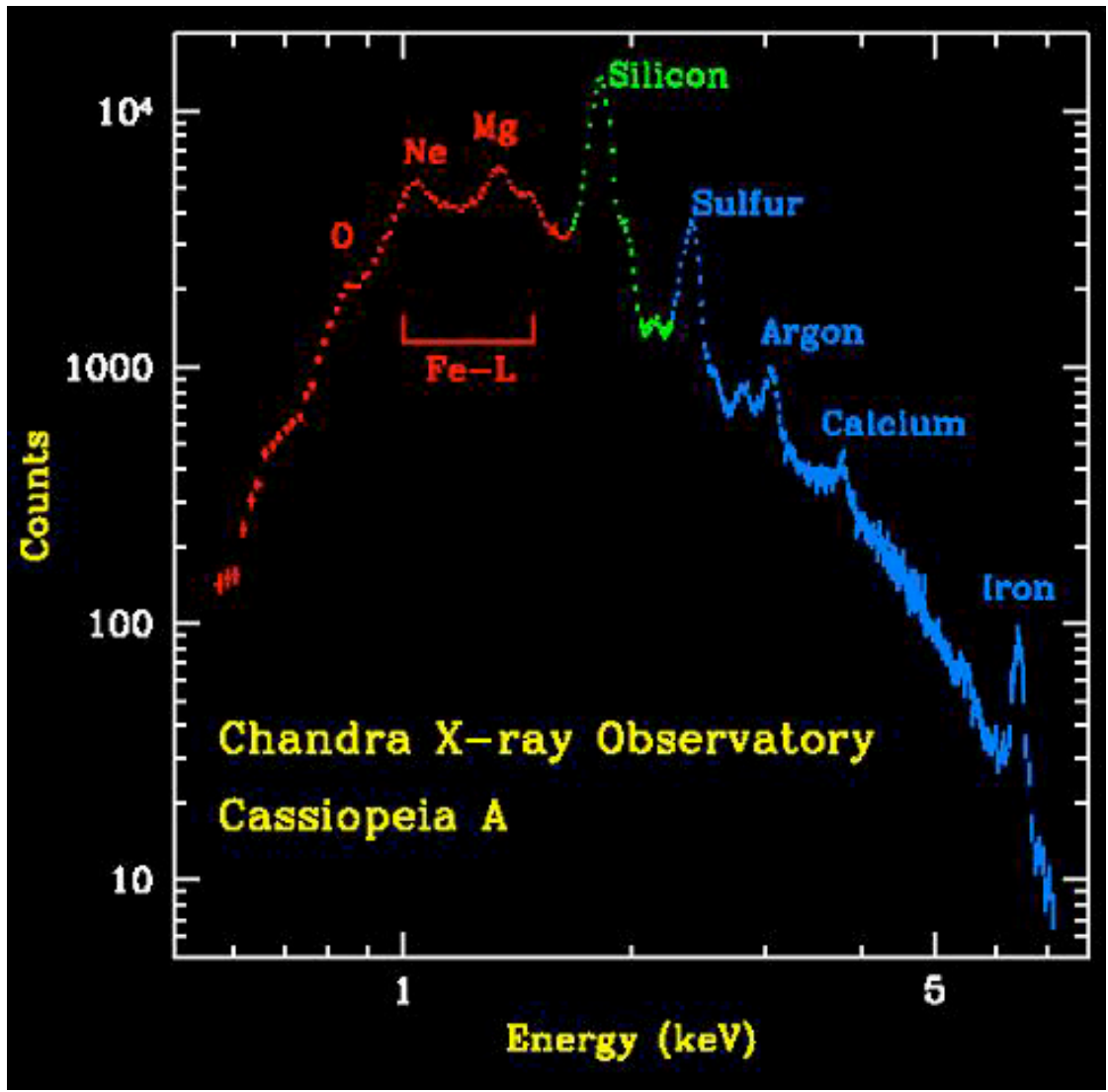

Fig.1. *Chandra* spectrum of the star Cassiopeia A, illustrating characteristic x-ray peaks.

## Applications of photoionization of atomic elements.

Space-based ultraviolet (UV) observations of emission and absorption lines from ions of carbon, nitrogen, oxygen, sulphur, and silicon in photoionized sources play an important role in addressing many of the astrophysical issues listed above. *Hubble Space Telescope* (*HST*) observations of carbon and sulphur ions in clouds have been used to study the origin and chemical evolution of gas at large distances from galaxies. Observations have provided a wealth of photon data on the ionizing spectrum and discriminated between an extragalactic radiation field due to AGN or O-type stars from starburst galaxies. The x-ray spectra of the outflowing gas in Seyfert galaxies have been measured by *Chandra*; absorption edges and dozens of x-ray emission lines are observed in their spectra.

## Selected types of atomic processes with photons.

The atomic processes occurring for a photon (light) with sufficient energy ($h\nu$, where $h$ is Planck's constant and $\nu$ is the frequency) to remove an outer- or inner-shell electron in a collision with a complex atom or ion containing a specific number of electrons are:

1. Valence (outer) shell single photoionization.

2. Photoexcitation (the mechanism of electron excitation by photon absorption, when the energy of the photon is too low to cause photoionization) and photoabsorption (**Fig. 2**).

3. Photoionization with photoexcitation.

4. Fluorescence (emission of a photon) from the resulting atomic ion (**Fig. 3**). The hole (vacancy in the innermost shell) is caused by the incoming photon having sufficient energy to remove it.

5. Auger processes (atomic transitions that do not emit light), where a secondary electron is ejected due to removal of an inner-shell electron. A photon is ejected by core relaxation of the subsequent ion. The process was named after Pierre Victor Auger.

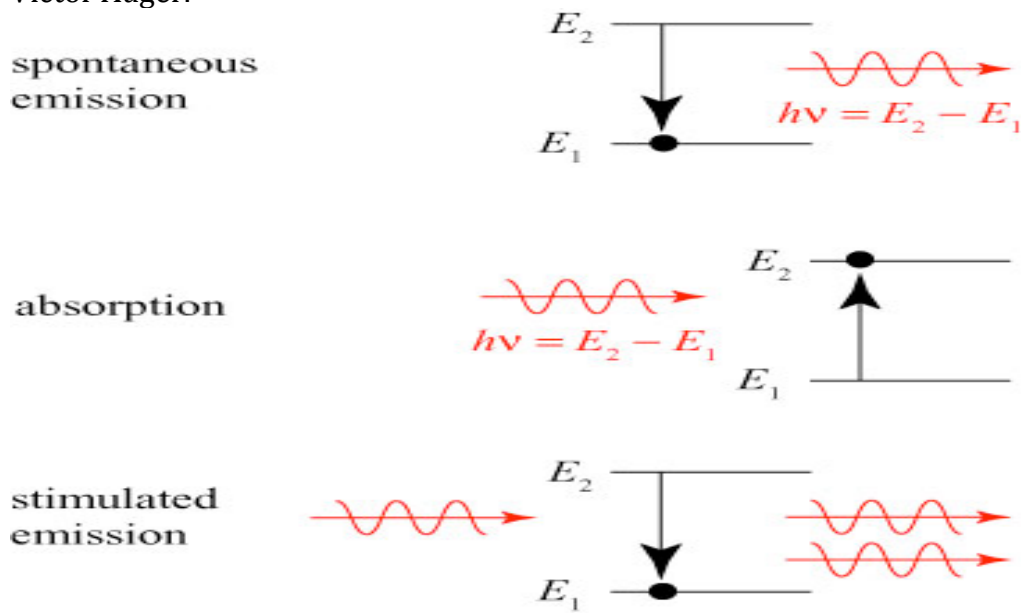

Fig. 2. Energy level diagrams of an atom illustrating the origin of line spectra. The absorption of a photon of energy $h\nu$ causes atomic transition from levels $E_1$ to $E_2$. Transition from level $E_2$ to $E_1$ causes emission of a photon.

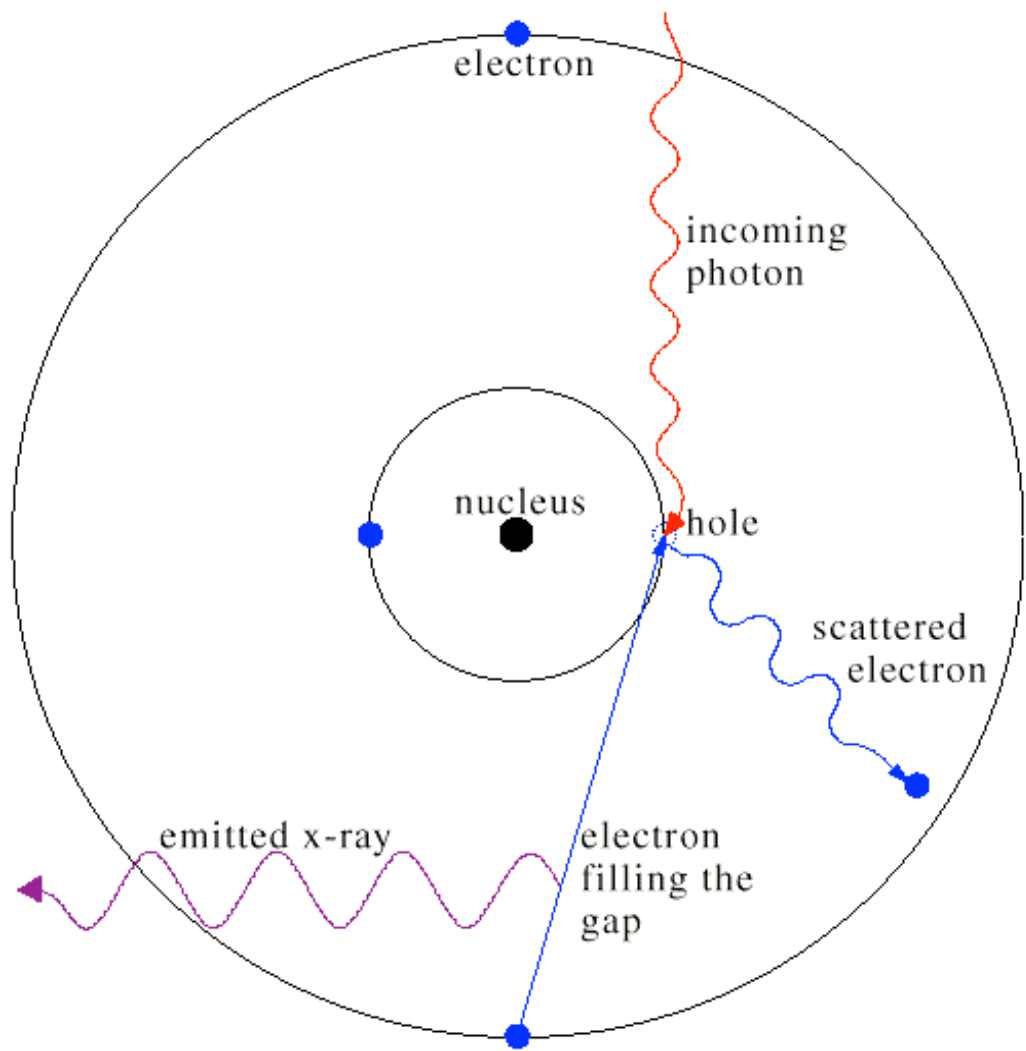

Fig. 3. A pictorial representation, of x-ray fluorescence in atomic systems. A hole is created by removal of an innermost electron by an incoming photon. An outer electron then fills this hole, causing the emission of a second photon, the emitted x-ray photon (illustrated by the purple photon trace), which is the x-ray fluorescence.

In the vacuum-ultraviolet wavelength region, the photoionization process is the dominant interaction between the photon and atoms (or electrons). In the soft x-ray region, that is, photon energies in the range 0.2–2 keV, or wavelengths in the region of approximately 0.5–4.5 nm, the Auger process dominates. Theoretically, sophisticated numerical methods are required to solve the quantum mechanical equations of motion governing the dynamics and interactions of the electron and photons to obtain the atomic collision observables. Continuous or line spectra may be obtained, the latter consisting of emission or absorption lines (**Fig.4**), for various wavelengths across the electromagnetic spectrum (**Fig. 5**).

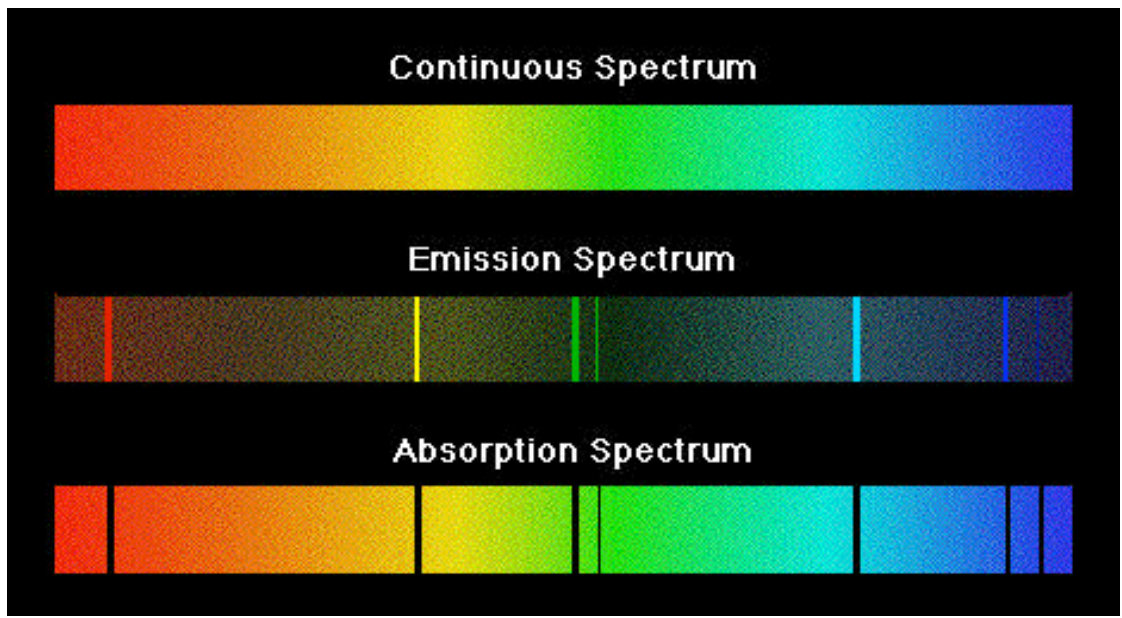

Fig. 4. Continuous spectrum and two types of line spectra

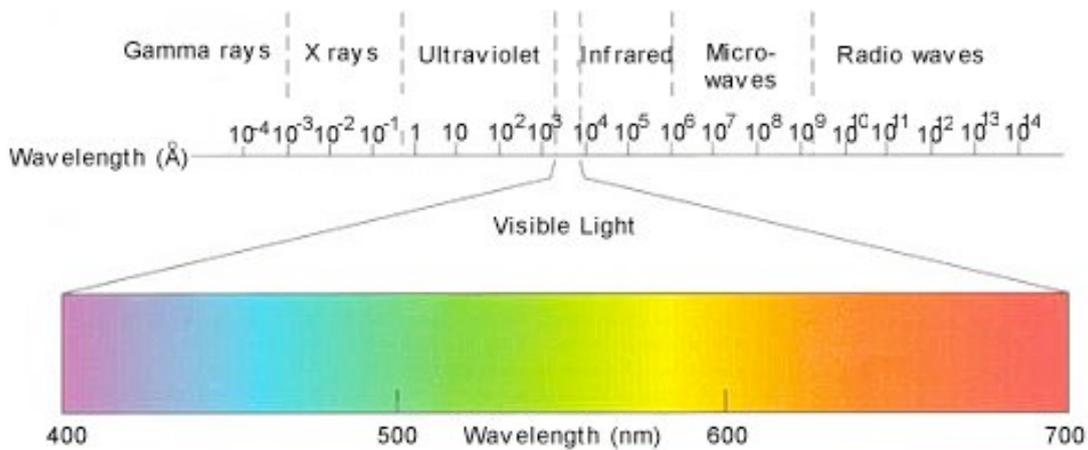

Fig. 5. The electromagnetic spectrum over a selected wavelength range.

## Photoionization of low-mass atomic elements.

The body of information concerning photoionization processes in atomic ions is mainly theoretical knowledge that requires benchmarking against experimental measurements from modern-day synchrotron facilities, such as the Advanced Light Source (ALS), SOLEIL, ASTRID, and PETRA, which provide electromagnetic radiation of high intensity and coherence. The motivation underlying these studies is to obtain a better understanding of the electron-electron interactions occurring in high-temperature environments, stellar atmospheres, and laboratory plasmas.

Knowledge of the photoionization cross section of the two-electron lithium ion provides an essential benchmark for future photoionization studies on more abundant highly ionized two-electron species such as carbon, nitrogen, oxygen, and neon. Such details are important in determining the mass of missing baryons (heavy subatomic particles made up of quarks, which are elementary particles and a fundamental constituent of matter) in the x-ray spectrum of the warm-hot intergalactic medium (WHIM).

Two-electron promotions in the lithium ion produce doubly excited states (states lying above their lowest energy levels) that can autoionize (undergo transitions that do not emit light) to form a single-electron ion and an outgoing free electron. The strongest process from the ground state of the lithium ion is the population of these doubly excited resonance states. Experimental studies on the lithium ion in the late 1970s resolved the spectrum in the wavelength region spanning 20–5 nm, observing strong peaks in the absorption spectrum that were attributed to these doubly excited states.

Two decades elapsed before the advent of third-generation synchrotron light sources produced high-resolution measurements on the lithium ion, first at the ALS and then at the Super-ACO facility. These experiments produced extremely high-quality data for several peaks in the spectrum of this system. This vast improvement in the magnitude of photon energy resolution yielded extremely accurate resonance energy positions, widths, and shapes of the resonances, providing stringent tests of theory (**Fig. 6**).

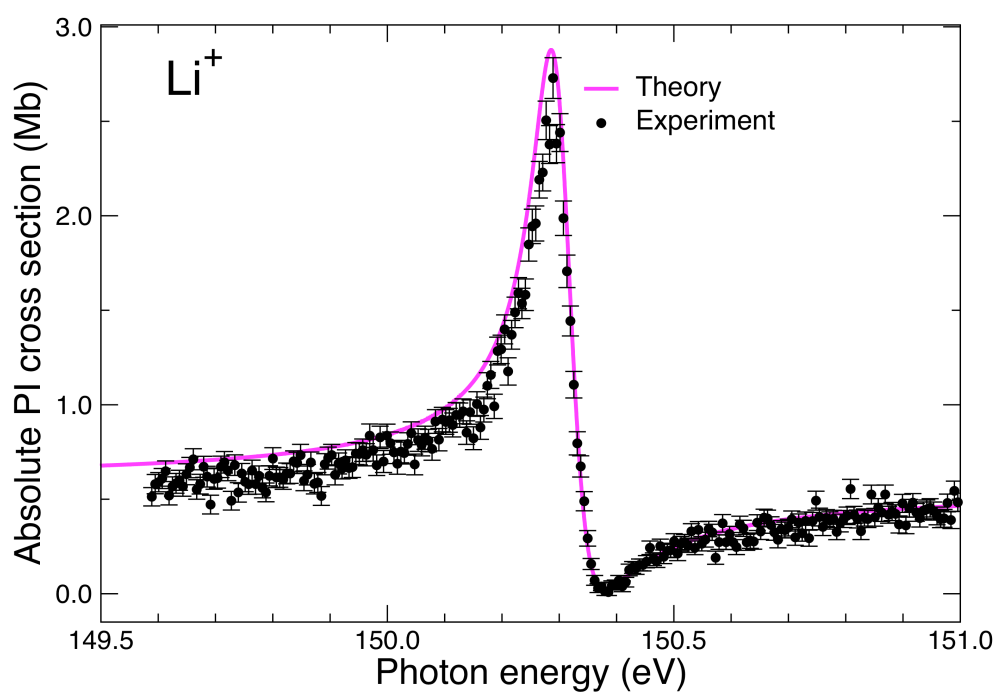

Fig. 6. Comparison of experimental and theoretical cross sections for photoionization (PI) of Li$^+$ ions in the photon energy range 149.5–151 eV. The measurements were obtained with an energy resolution of 38 meV at the Advanced Light Source (ALS). (*Adapted from S. W. J. Scully et al., Doubly excited*

*resonances in the photoionization spectrum of Li⁺: experiment and theory, J. Phys. B.: At. Mol. Opt. Phys., 39:3957–3968, 2006*)

Additional information such as the lifetime of atomic states may be determined from sophisticated numerical methods in quantum physics. Line fluorescence yields ω (emission of light by the system that has absorbed it) may also be determined, assuming statistical population of the states. The fluorescence yield ω, together with branching ratios, serves as a quantitative measure of the decay mode of the atomic state.

## Photoionization of atomic elements of intermediate mass.

For atomic elements in the periodic table of intermediate charge value $Z$, relativistic effects become increasingly important. For comparison with high-resolution measurements from synchrotron radiation facilities, state-of-the-art numerical methods that include semi-relativistic effects are necessary. This optimizes experimental beam time, as coarse energy scans may be used in resonance-free regions and fine scans at high resolution in energy ranges with dense resonance structure, as illustrated in recent work on Ar⁺ ions at the ALS.

The interaction of photons with atomic ions is an important process in determining the ionization balance (the population ratio related to atomic collision processes) of the abundances of elements in photoionized astrophysical nebulae. In nebulae (interstellar clouds of gas and dust, containing hydrogen, helium, and other atoms and molecules, some of which are ionized), the formations of gas, dust, and other materials "clump" together to form larger masses, which attract further matter, and eventually will become massive enough to form stars.

Among atomic elements heavier than iron ($Z$ = 26), known as trans-iron elements, neutron ($n$)-capture elements (for example, selenium, krypton, bromine, xenon, rubidium, barium, and lead) have been detected in a large number of ionized nebulae. Elements are produced by slow or rapid neutron-capture nucleosynthesis (the process of creating new atomic nuclei from preexisting nucleons). With the formation of stars, heavier nuclei were created from hydrogen and helium by stellar nucleosynthesis, a process that continues today. Some elements, those lighter than iron, are delivered to the interstellar medium (the matter that exists in the space between the star systems in a galaxy) in the last stages of evolution of dying low-mass stars in the nonexplosive ejection of the outer envelope gases of planetary nebulae before these stars continue to form white dwarfs (a small star composed of electron-degenerate material that no longer undergoes fusion reactions). Measuring the abundances of these elements helps to reveal their dominant production sites in the universe, and details of stellar structure, mixing, and nucleosynthesis.

The level of enrichment of individual elements is strongly sensitive to the physical conditions in the stellar interior. Uncertainties in the photoionization and recombination (the formation of neutral atoms from the capture of free electrons by the cations in a plasma) cross-section data of $n$-capture element ions can result in elemental abundance uncertainties of a factor of two or more.

## Photoionization of atomic elements of heavy mass.

For atoms or ions with charge greater than 30, relativistic and electron correlation effects are essential to include, to obtain accurate modeling parameters. Computationally, the number of atomic levels and coupled states involved dramatically increases.

To address the challenge of electrons or photons interacting with heavy atomic systems containing hundreds of levels resulting in thousands of scattering channels, efficient parallel numerical algorithms were developed. This allows the single-photon ionization modeling of trans-iron elements and opens the doorway to comprehensive large-scale calculations along isonuclear sequences (atomic systems comprising of the same nucleus but in successive stages of ionization, for example, $Se^+$, $Se^{2+}$, $Se^{3+}$,…).

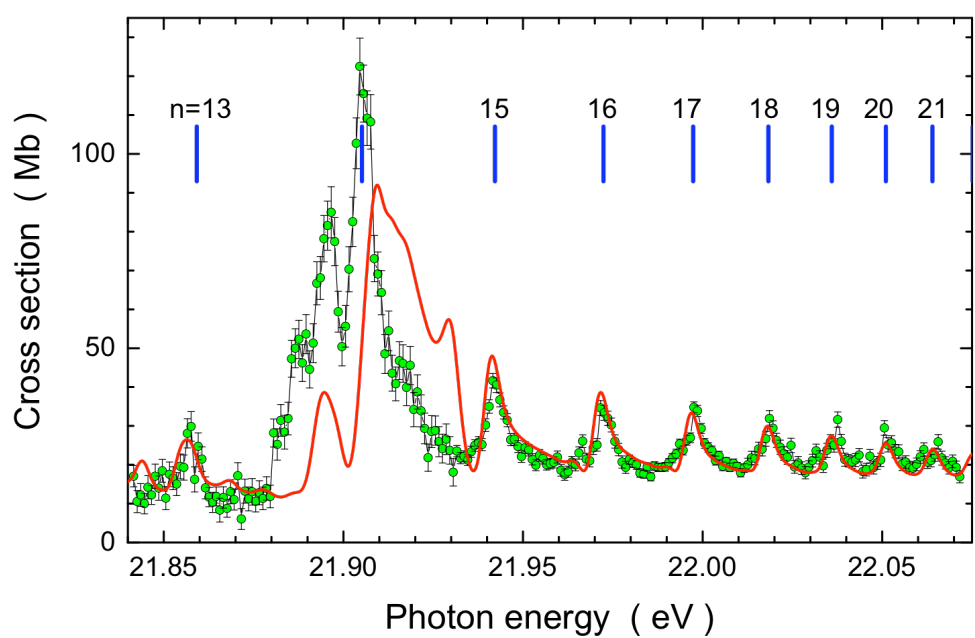

Fig. 7. Advanced Light Source (ALS) $Xe^+$ photoionization experimental cross-section data (green circles) for photon energies in the range 21.84–22.08 eV, at a photon energy resolution of 4 meV. Theoretical results (red line) at 4 meV include fully relativistic effects. The bars mark the energies of the $5p \rightarrow nd$ resonances in the $Xe^+$ spectra. (*Courtesy of B. M. McLaughlin and C. P. Balance, Photoionization cross section calculations for the halogen-like ions $Kr^+$ and $Xe^+$, J. Phys. B: At. Mol. Opt. Phys., 45:085701, 2012*)

Selenium ions were chosen for study because they have been detected in nearly twice as many planetary nebulae (70 in total) as any other trans-iron element. Experimental photoionization cross-section data results for singly-ionized selenium obtained at the ALS in the near-threshold energy region show excellent agreement with theoretical results obtained from fully relativistic calculations.

Xenon and krypton ions are of importance in human-made plasmas such as light sources for semiconductor lithography (a process used in the electronic industry to selectively remove parts of a thin film), ion thrusters for spacecraft pro-pulsion, and nuclear fusion plasmas. Xenon and krypton ions have also been detected in cosmic objects, planetary nebulae, and the ejected envelopes of low- and intermediate-mass stars. For an understanding of these plasmas, accurate cross sections are required for ionization and recombination processes that govern the charge balance of ions in plasmas. Krypton and xenon ions are also of importance in tokomak (a device that uses a magnetic field to confine a plasma in the shape of torus) plasmas for fusion physics.

Photoionization experimental results from the ALS synchrotron radiation facility for singly-charged xenon ions in the near-threshold region show suitable agreement with theoretical results including fully relativistic effects (**Fig 7**).

## X-ray spectroscopy of atomic elements.

X-rays are short-wavelength electromagnetic radiation produced by the deceleration of high-energy electrons or by electronic transitions in the inner orbitals of atoms. The wavelength range of x-rays is approximately $10^{-15}$–$10^{-8}$ m; conventional x-ray spectroscopy is largely confined to the region of about $10^{-11}$–$2.5 \times 10^{-9}$ m. X-ray spectroscopy is a form of optical spectroscopy that utilizes emission, absorption, scattering, fluorescence, and diffraction of x-rays.

Inner-shell (x-ray) photoionization processes play important roles in many physical systems, including a broad range of astrophysical objects as diverse as quasars, the atmospheres of hot stars, nebulae, novae, and supernovae. Inner-shell studies of carbon and its ions indicate that high-quality theoretical data are required to model the observations in the x-ray spectrum of the bright blazar Makarian 421, observed by the *Chandra X-ray Observatory* (**Fig. 8**). Abundances for carbon, nitrogen, and oxygen in their various ionized stages are essential for photoionization models applied to the plasma modeling in a variety of planetary nebulae. Nitrogen abundance in particular plays a fundamental role in studies of planetary nebulae because it is a key tracer of the carbon-nitrogen-oxygen processing.

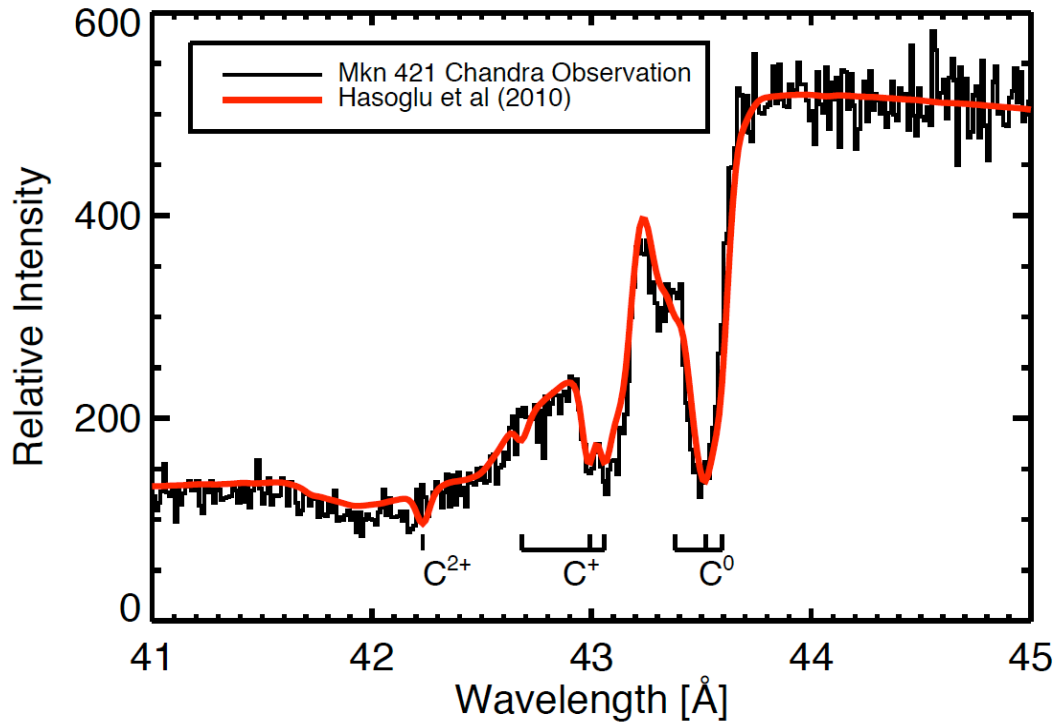

Fig. 8. Modeling of the observed x-ray spectrum of the bright extragalactic x-ray source blazar Makarian 421 near the carbon *K*-edge, obtained from the *Chandra X-ray Observatory*. The series of connected black bars are the observed spectrum. The blue and red curves are modeling of the observations by two different groups. Wavelengths of transitions of carbon and its ions are also indicated. (*Courtesy of M. F. Hasoglu et al., K-shell photoabsorption studies of the carbon isonuclear sequence, Astrophys. J., 724:1296-1304, 2010*)

Inner-shell photoionization of atomic nitrogen is of crucial importance to the energetics of the terrestrial upper atmosphere and, together with photoionization of atomic oxygen and molecular nitrogen (which is the most abundant species), determines the ion-neutral chemistry and temperature structure of the upper atmosphere, ultimately through the production of nitric oxide (NO). The NO abundance is highly correlated with the soft x-ray irradiance (the power per unit area radiated by a surface), but uncertainties in the photoionization cross sections and solar fluxes remain. Highly accurate experimental results from the ALS for the photoionization cross section of atomic nitrogen in the vicinity of the *K*-edge, where atomic resonances features (peaks) are present in the spectrum, are well reproduced by theory (**Fig.9**).

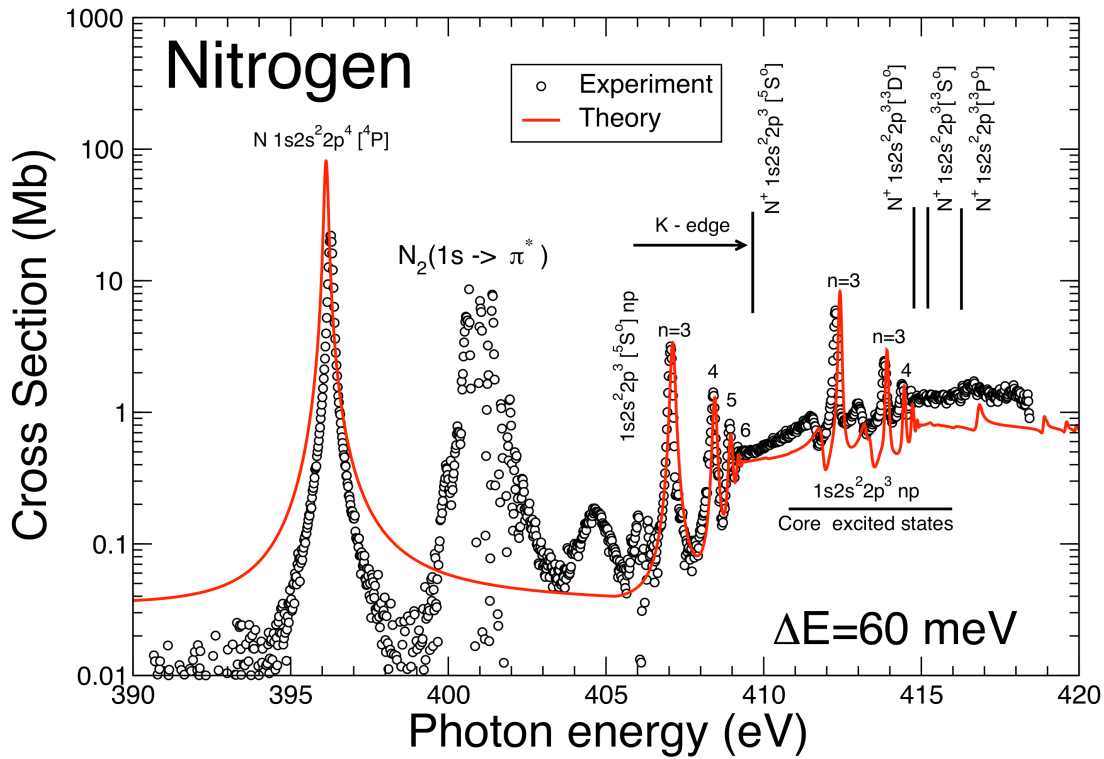

Fig. 9. Atomic nitrogen, photoionization cross sections from the Advanced Light Source (ALS) compared with theory. Theoretical results are convoluted with a 60-meV FWHM (full width at half maximum) Gaussian to simulate experiment. Experimental results include additional molecular components between 400 eV and 406 eV. Spectroscopic notation is to identify the molecular and atomic states of the peaks in the spectra. (*Courtesy of M. M. Sant'Anna et al., K-shell x-ray spectroscopy of atomic nitrogen, Phys. Rev., Lett., 107:033001, 2011*)

PETRA III, in Hamburg, Germany, which began operation in 2009, is the most brilliant storage-ring-based x-ray source in the world. This exceptionally high brilliance offers scientists outstanding experimental opportunities. PETRA III benefits researchers investigating very small samples or those requiring tightly collimated and very short-wavelength x-rays for their experiments. Investigations on the resonant fluorescence from highly charged ions of iron have been studied at energies of 6–7 keV, with signals from 2–6-electron iron ions being observed. Further studies plan to look at photon impact on these highly charged iron ions in the *K*-excitation channel (the case where one of the innermost electrons is excited) with subsequent autoionization.

## Conclusions.

In the past decade or more, observations by orbiting satellites coupled with major experimental advances from light sources have made it possible to determine accurate results for cross sections of photons interacting with atomic systems in their neutral, singly ionized, or highly ionized states. High-resolution cross-section measurements at third-generation light sources have pushed the boundaries of theoretical methods.

Major advances in state-of-the-art theoretical methods coupled with advances in computational architectures have opened the doorway to large-scale computations on elements across the periodic table, allowing the inclusion of fully relativistic effects. Various stages of ionization, necessary for the many applications in astrophysics, can be studied in the absence of experimental values. We therefore look forward to another decade of major advances in experiment and theory and the synergistic relationship between them.

For background information see ASTRONOMICAL SPECTROSCOPY; ATOMIC STRUCTURE AND SPECTRA; AUGER ELECTRON SPECTROSCOPY; AURORA; CARBON-NITROGEN-OXYGEN CYCLES; CHANDRA X-RAY OBSERVATORY; FLUORESCENCE; GALAXY, EXTERNAL; HUBBLE SPACE TELESCOPE; INTERSTELLAR MATTER; LIGHT; NEBULA; NUCLEOSYNTHESIS; PHOTOIONIZATION; PHOTON; PLANETARY NEBULA; PLASMA (PHYSICS); SYNCHROTRON RADIATION; ULTRAVIOLET ASTRONOMY; X-RAY ASTRONOMY; X-RAY SPECTROMETRY; X-RAY TELESCOPE; X-RAYS in the McGraw-Hill Encyclopedia of Science & Technology.

Brendan M. McLaughlin; Connor P. Ballance

## Bibliography.

**URLs**

Chandra X-ray Observatory

http://chandra.harvard.edu

NIST, Physical Measurement Laboratory

http://www.nist.gov/pml/data

Advanced Light Source

http://www-als.lbl.gov